\documentclass[pra,twocolumn,superscriptaddress,amsmath,amssymb]{revtex4-1}
\usepackage[usenames,dvipsnames]{color}
\usepackage{graphicx,mathptmx,bm,epstopdf,wasysym,subfigure,lettrine,ragged2e,anyfontsize,hyperref}

\newcommand{\tr}[1]{\textnormal{tr}{\left\{#1\right\}}}
\newcommand{\SP}[1]{\textnormal{Sp}{\left\{#1\right\}}}
\newcommand{\TP}[1]{{#1}^\mathrm{\,\textsc{t}}}
\newcommand{\D}{\mathrm{d}}
\newcommand{\I}{\mathrm{i}}

\newcommand{\rvec}[1]{\bm{#1}}
\newcommand{\dyadic}[1]{\bm{#1}}

\newcommand{\GMAT}{\pmb{G}}
\newcommand{\DET}[1]{\det\!\left\{#1\right\}}

\begin{document}

\title{Surmounting intrinsic quantum-measurement uncertainties in Gaussian-state tomography with quadrature squeezing}

\author{Jaroslav {\v R}eh{\'a}{\v c}ek}
\email{rehacek@optics.upol.cz}
\affiliation{Department of Optics, Palack{\'y} University, 17. listopadu 12, 77146 Olomouc, Czech Republic}

\author{Yong Siah Teo*}
\email[Correspondence and requests for materials should be addressed to Y.~S.~T. e-mail: ]{yong.teo@upol.cz}
\affiliation{Department of Optics, Palack{\'y} University, 17. listopadu 12, 77146 Olomouc, Czech Republic}

\author{Zden{\v e}k Hradil}
\email{hradil@optics.upol.cz}
\affiliation{Department of Optics, Palack{\'y} University, 17. listopadu 12, 77146 Olomouc, Czech Republic}

\author{Sascha Wallentowitz}
\email{swallent@fis.puc.cl}
\affiliation{Instituto de F{\'i}sica, Pontificia Universidad Cat{\'o}lica de Chile, Casilla 306, Santiago 22, Chile}

\begin{abstract}
We reveal that quadrature squeezing can result in significantly better quantum-estimation performance with quantum heterodyne detection (of H.~P.~Yuen and J.~H.~Shapiro) as compared to quantum homodyne detection for Gaussian states, which touches an important aspect in the foundational understanding of these two schemes. Taking single-mode Gaussian states as examples, we show analytically that the competition between the errors incurred during tomogram processing in homodyne detection and the Arthurs-Kelly uncertainties arising from simultaneous incompatible quadrature measurements in heterodyne detection can often lead to the latter giving more accurate estimates. This observation is also partly a manifestation of a fundamental relationship between the respective data uncertainties for the two schemes. In this sense, quadrature squeezing can be used to overcome intrinsic quantum-measurement uncertainties in heterodyne detection.
\end{abstract}

\date{\today}

\begin{widetext}
\maketitle
\end{widetext}
\small
Continuous-variable (CV) quantum tomography in phase space, or the diagnostics and reconstruction
 of observables of infinite Hilbert-space dimension in the continuous phase-space representation,
 is an indispensable technique for characterizing quantum light sources. These sources play a crucial
 role in practical optical quantum cryptography protocols \cite{cvQC1,cvQC2,cvQC3,cvQC4,cvQC5},
 many of which involve security analysis on Gaussian quantum states
 \cite{gaussQC1,gaussQC2,gaussQC3,gaussQC4,gaussQC5} --- namely the coherent and squeezed states.
 Not only do Gaussian sources have properties that are intimately related to optomechanical phenomena \cite{optomech1,optomech2} and extensively exploited in quantum metrology \cite{qmetro1,qmetro2}, but they are also considered in the study of protocols such as state teleportation \cite{cvtelep1,cvtelep2}, dense coding
 \cite{cvdc1,cvdc2} and cloning \cite{cvclone1,cvclone2}. The list of references is certainly
 non-exhaustive, and the reader is referred to two review articles on the subject of CV quantum
 optical communications \cite{revCVQI1,revCVQI2}.


\begin{figure*}[t]
\centering
\subfigure[~Homodyne detection]{\includegraphics[width=0.45\textwidth]{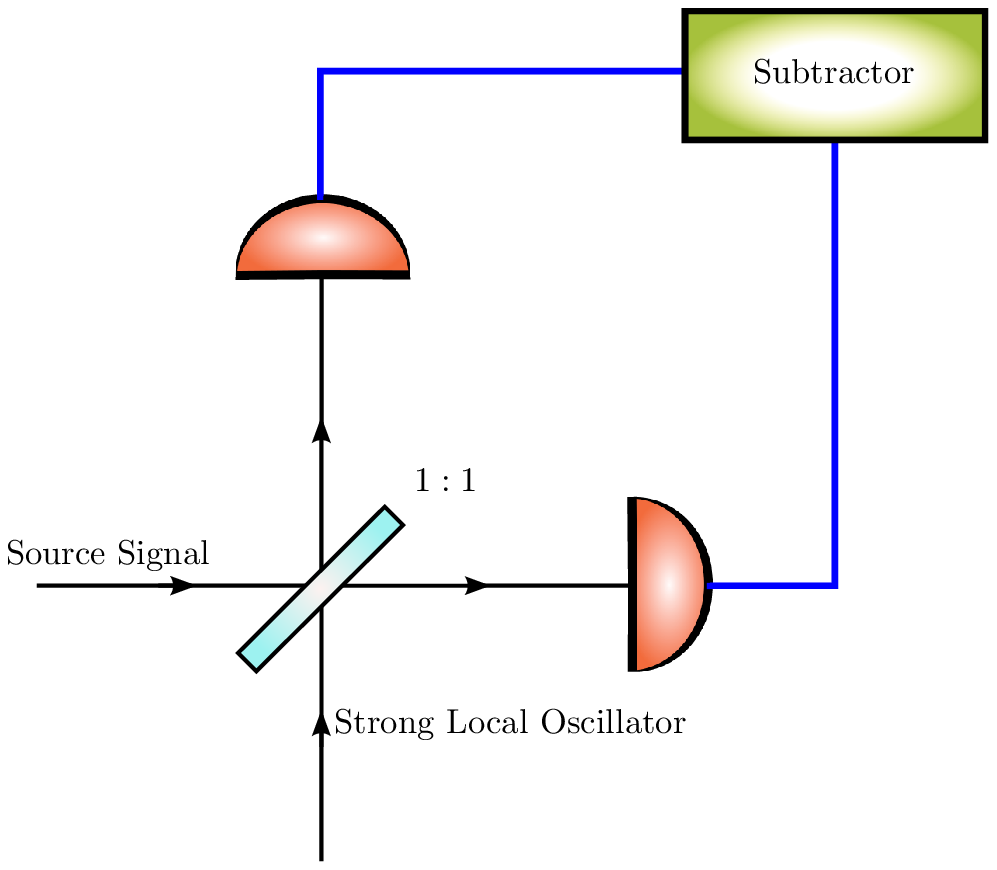}}
\subfigure[~Heterodyne detection]{\includegraphics[width=0.45\textwidth]{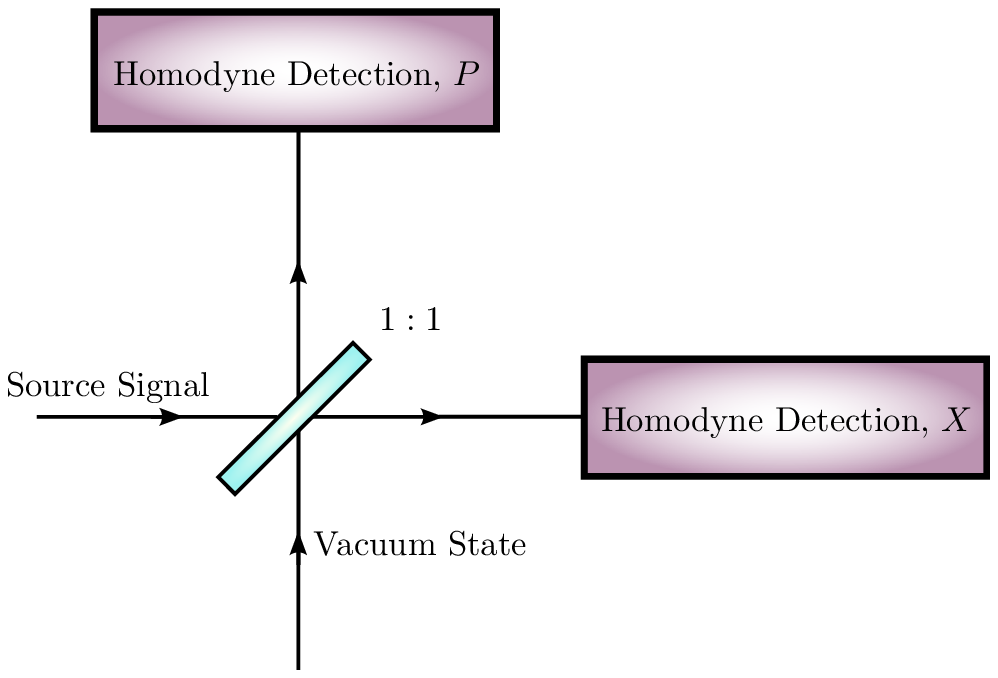}}
\caption{\textbf{Schematics for homodyne and heterodyne detections.}~In our context, the heterodyne scheme shall be understood as a double-homodyne scheme (shown in (b)) that measures two complementary sets of quadrature eigenstates (those of position $X$ and momentum $P$) simultaneously.}
\label{fig:fig1}
\end{figure*}

It therefore goes without saying that quantum tomography techniques for quantum states of light
 are of major interest in recent years. One of the most popular techniques, quantum homodyne detection
\cite{homdet1,homdet2,homdet3}, is used in quantum optics to measure intensities of light signals
 from the outputs of a beam splitter that coherently merges the light mode from the source and
 that from a local oscillator, or reference coherent state. The result is an approximate
 measurement of the eigenstates of a rotated photonic quadrature whose phase angle depends
on the phase of the local oscillator. The homodyne data obtained per binned angle constitute a
distribution of points along a particular phase-space cut defined by this angle. This
distribution is precisely the marginal distribution of the Wigner function of the quantum
 state of the source over the complementary quadrature. There is another well-known optical technique,
 heterodyne detection \cite{hetdet1,hetdet2,hetdet3,hetdet4,hetdet5,hetdet6,hetdet7,hetdet8}, which has been extensively used to simultaneously probe a pair of optical beams of different frequencies in order to measure their relative phase. Compared to quantum homodyne detection, there are apparently relatively fewer published works on using its quantum variant to perform quantum tomography. This involves simultaneously measuring signal intensities of beams that are split from a single
source signal by a beam splitter, thus realizing
 the approximate measurement of two rotated quadratures (position and momentum say) that are
 complementary to each other (see Fig.~\ref{fig:fig1} for schematics). Because of the nature of such a quantum measurement, we shall understand the heterodyne scheme discussed here as a quantum double-homodyne tomography scheme, and henceforth, with common understanding, drop the adjective ``quantum'' when referring to these schemes. The heterodyne data obtained constitute a distribution of points according to the Husimi~Q function of the state.

Manipulation of quantum tomograms have been carried out to address certain scalar quantities of interest (purity \cite{bellini} for instance). Throughout the discussion, we shall
 focus on the reconstruction of covariance matrices that fully characterize the Wigner function of single-mode Gaussian states using the two CV schemes.

There exist elements that intrinsically affect the tomographic accuracy of quantum-state
 (or observable) reconstruction with typical CV tomography data. Two kinds of data disturbances
 that are ubiquitous in every experiment are statistical errors owing to the reconstruction
 with only a finite number of data points collected, as well as instrumental errors arising from
 errors in detector efficiency calibrations, measurement settings, \emph{etc.}. There are other
 tomographic elements that are particular to the two schemes of interest. In homodyne tomography,
 the phase space is sampled in terms of cuts resulted from marginalizations of the Wigner function
 over the complementary quadratures relative to the measured ones. The unknown quantum state,
 or any other field observable for that matter, is reconstructed by post-processing the homodyne
 data --- also known as tomograms for historical reasons \cite{CT} --- to reverse the marginalization
 in order to recover the full state. The heterodyne data, on the other hand, already consists
 of sample points distributed from the Q~function in phase space. Nevertheless, it is well-known
 that heterodyne tomography gives quantum measurement errors corresponding to the Arthurs-Kelly
 uncertainty relation, which originates from the simultaneous measurement of complementary
 quadrature observables \cite{complem_meas1,complem_meas2}.

Two main goals are achieved in this discussion. Firstly, we reveal that for data collected along any pre-chosen phase-space direction, the data uncertainties for heterodyne detection typically become smaller relative to those for homodyne detection for highly-squeezed quantum states. This, as it turns out, is a result of an intimate relationship between the conditional variance of heterodyne data and the marginal variance of homodyne data, which defines the underlying statistical behavior for such data. In the context of covariance estimation, where covariances are the statistical quantities that completely characterize all Gaussian states, the difference in tomographic performance for covariance-matrix reconstruction between the two detection schemes is a sophisticated consequence of this fundamental relationship. This brings us to the second goal --- the investigation of the optimal tomographic accuracy of unbiased covariance-matrix reconstruction for these two CV schemes. To appreciate the underlying physical framework, it will be shown that if the Arthurs-Kelly-type errors were absent, heterodyne detection on single-mode Gaussian states indeed always outperforms homodyne detection as far as the optimal limit of unbiased covariance-matrix reconstruction is concerned. In real scenarios where these errors are always present, this article shall demonstrate that heterodyne detection can still beat homodyne detection in optimal tomographic accuracy in many situations of practical relevance.

\bigskip
\noindent{\normalsize{\textbf{Results}}}
\small

\noindent{\textbf{Notations.}}~The covariance matrix that characterizes the Wigner function of a
 single-mode Gaussian quantum state can be represented by the positive $2\times2$ real matrix
\begin{equation}
\dyadic{G}_\textsc{w}=\dfrac{1}{2}\begin{pmatrix}
2\left<(\Delta X)^2\right> & \left<\{\Delta X,\Delta P\}\right>\\
\left<\{\Delta X,\Delta P\}\right> & 2\left<(\Delta P)^2\right>
\end{pmatrix}\geq\dfrac{1}{2}\begin{pmatrix}
0 & -\I \\
\I & 0
\end{pmatrix}
\end{equation}
involving expectation values of functions of the respective position- and momentum-quadrature
 standard deviations $\Delta X$
 and $\Delta P$. The matrix elements of $\dyadic{G}_\textsc{w}$ specify the size, shape, and
orientation of the ellipsoidal level contours of the Wigner function.\\

\noindent{\textbf{Data uncertainties along a phase-space direction.}}~In this section, we shall perform a generic analysis on uncertainty regions for the two schemes. To begin, we recall two different notions of probability distribution functions from statistics that can be derived from a joint probability distribution function of a set of parameters taking random values. The first kind of distribution functions, the \emph{conditional distribution function}, is a probability distribution function, which values are defined by a subset of these parameters after fixing the values for the rest of the parameters. The second kind, the \emph{marginal distribution function}, is a probability distribution function of a subset of these parameters by taking an average over the rest of the parameters. Data sampling according to these two kinds of distributions respectively produce data uncertainties that are quite different. The data uncertainty that goes with the marginal distribution are never less than that which goes with the conditional distribution.

\begin{figure}[h!]
\centerline{
\includegraphics[width=\columnwidth]{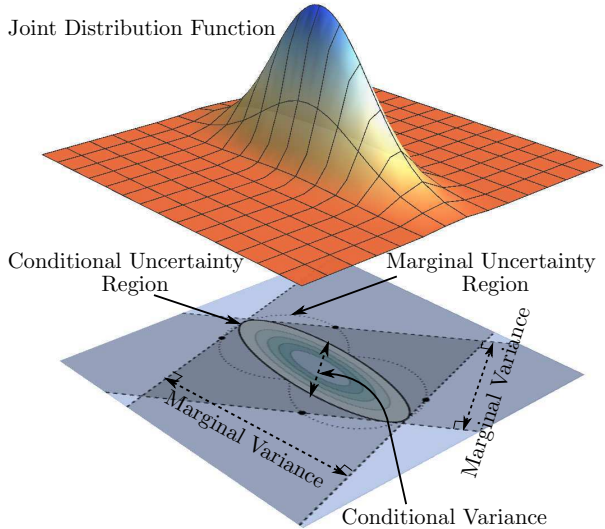}}
\caption{\textbf{Marginal and conditional uncertainties of a joint distribution function.}~For a joint distribution function over a two-variable (two-dimensional) space, the marginal variance for data sampled along a particular reference direction (indicated by a double-headed arrow) can be understood as a shadow cast from the joint uncertainty region in the orthogonal direction (indicated by the corresponding perpendicular pair of dashed lines). On the other hand, the conditional variance is directly obtained by slicing the joint uncertainty region about its center along the same reference direction. The marginal uncertainty region is the region defined by pairs of points bounding the marginal variances in all directions, which is similar to how the conditional uncertainty region is related to the conditional variances.}
\label{fig:marg_cond}
\end{figure}

Figure~\ref{fig:marg_cond} illustrates this point for a Gaussian joint distribution function and the proof is straightforward. Let us consider a Gaussian conditional probability distribution of two random variables characterized by a two-dimensional covariance matrix $\dyadic{G}$, and denote the marginal variance by $\sigma^2_\theta$, and the conditional variance by $\Sigma^2_\theta$. We shall fix the reference direction, defined by the angle $\theta$ and shown graphically as dashed straight lines in Figure~\ref{fig:marg_cond}, along which we choose to either average/integrate over (marginal variance) or slice (conditional variance), to be parallel to the basis vector $\rvec{u}_\theta\,\widehat{=}\,(\cos\theta\,\,\,\sin\theta)$.

Following the usual definitions of the two variances for a Gaussian distribution, we have
\begin{equation}
\sigma^2_\theta=\TP{\rvec{u}}_\theta\,\dyadic{G}\,\rvec{u}_\theta\equiv G_{uu,\theta}
\end{equation}
and
\begin{align}
\Sigma^2_\theta&=\big(\TP{\rvec{u}}_\theta\,\dyadic{G}^{-1}\,\rvec{u}_\theta\big)^{-1}\nonumber\\
&=\frac{G_{uu,\theta}G_{vv,\theta}-G_{uv,\theta}^2}{G_{vv,\theta}}\le G_{uu,\theta}=\sigma^2_\theta\,.
\end{align}
The simple inequality implies that the uncertainty for data acquired from the marginal distribution is always greater than those acquired from the full conditional distribution. The equality holds only in the following two cases: (i) The integration, or shadow projection, is done along principal axes of the ellipse; (ii) The ellipse is actually a circle (zero eccentricity) so that all directions are principal.

The preceding background introduction sets the stage for clarifying an important operational difference between homodyne and heterodyne detection. In a homodyne measurement, the data obtained for a fixed angle $\theta$ is sampled according to the marginal distribution of the Wigner function. On the other hand, data gathered from a heterodyne measurement \emph{for the same angle} are sampled directly from the Q~function, which is the conditional distribution for the standard pair of complementary quadrature variables. Owing to the additional vacuum contribution present in the heterodyne measurement, the conditional covariance of the Q~function is larger than that of the Wigner function. As a consequence, the heterodyne-data uncertainty region is not always smaller than the homodyne-data uncertainty region. Nevertheless, for Gaussian states with highly elongated Wigner functions, the uncertainty region of the heterodyne data is indeed smaller than that of the homodyne data. In other words, if one picks any phase-space direction to collect both homodyne and heterodyne data originating from a source described by a highly elongated Wigner function, then on average, the heterodyne data have a smaller spread than the homodyne data. Details of the uncertainty region analysis are given in the \textbf{Methods} section under \textbf{Uncertainty regions.}.

Hence apparently, heterodyne detection is the better scheme for Gaussian states with highly elongated Wigner functions, and that profile asymmetry seems to be the crucial factor for data collected along a particular phase-space direction. However, as we shall soon witness, the two different statistical nature of the CV measurement schemes, as discussed previously, entangle with other data-processing factors in the schemes in a much more complicated way in quantum-state estimation.

To quantify the optimal tomographic performance between the two CV schemes, one needs to directly investigate the uncertainties of relevant tomographic quantities of interest, and measures of optimal tomographic accuracies generally depend on aspects of these uncertainty regions in a highly convoluted manner. Regardless, behind this underlying complexity lies a similarly elegant conclusion: Quadrature squeezing improves the tomographic performance of heterodyne data over homodyne data, thereby surmounting the intrinsic Arthurs-Kelly measurement uncertainties. That is, for highly-squeezed single-mode Gaussian states, heterodyne detection almost always gives a better tomographic performance than homodyne detection.\\

\noindent{\textbf{Covariance estimation - main result.}}~To formally compare
the optimal tomographic accuracy of any unbiased estimation of the covariance matrix, we adopt the
well-known tomographic measure $\mathcal{H}=\SP{\dyadic{F}^{-1}}$ --- the matrix
trace of the inverse of the Fisher information matrix $\dyadic{F}$ that is scaled with the number of copies $N$ --- that sets the Cram{\'e}r-Rao bound for all unbiased matrix estimators \cite{cramer-rao1,cramer-rao2}, the best possible tomographic performance any unbiased estimator can achieve. By denoting the bounds for homodyne and heterodyne detection respectively
by $\mathcal{H}_\textsc{hom}$ and $\mathcal{H}_\textsc{het}$, and the detector efficiency that is common
to both schemes by $\eta\leq1$, the results are summarized by the following closed-form expressions:
\begin{align}
\mathcal{H}_\textsc{hom}&=2\,\SP{\dyadic{G}_\textsc{hom}}\left(\SP{\dyadic{G}_\textsc{hom}}
+3\sqrt{\DET{\dyadic{G}_\textsc{hom}}}\right)\,,\nonumber\\
\mathcal{H}_\textsc{het}&=2\big[(\SP{\dyadic{G_\textsc{het}}})^2-\DET{\dyadic{G}_\textsc{het}}\big]\,;
\nonumber\\
\dyadic{G}_\textsc{hom}&=\dyadic{G}_\textsc{w}+\delta^{\textsc{(hom)}}_\eta\dyadic{1}\,,
\quad\delta^{\textsc{(hom)}}_\eta=\dfrac{1-\eta}{2\eta}\,,\nonumber\\
\dyadic{G}_\textsc{het}&=\dyadic{G}_\textsc{w}+\delta^{\textsc{(het)}}_\eta\dyadic{1}\,,
\quad\delta^{\textsc{(het)}}_\eta=\dfrac{2-\eta}{2\eta}\,.
\label{eq:hom_het}
\end{align}

Since the measure $\mathcal{H}$ is invariant under a rotation of $\dyadic{G}_\textsc{w}$, as it should,
we may parametrize this matrix
\begin{equation}
\dyadic{G}_\textsc{w}=\dfrac{\mu}{2}\begin{pmatrix}
1/\lambda & 0\\
0 & \lambda
\end{pmatrix}
\end{equation}
 using only two relevant parameters: $\mu\geq1$ that controls the size (related to the temperature) and $\lambda\geq1$ that controls
 the shape or the ratio of the major to the minor axis (related to the extent of quadrature squeezing). After establishing Eq.~\eqref{eq:hom_het}, all
 subsequent comparisons shall revolve around these equations. Since we are concerned with only profile
 estimation, the center of the Gaussian Wigner function, that is its mean value, is neglected.\\

\begin{figure*}[t]
\centering
\includegraphics[width=1\textwidth]{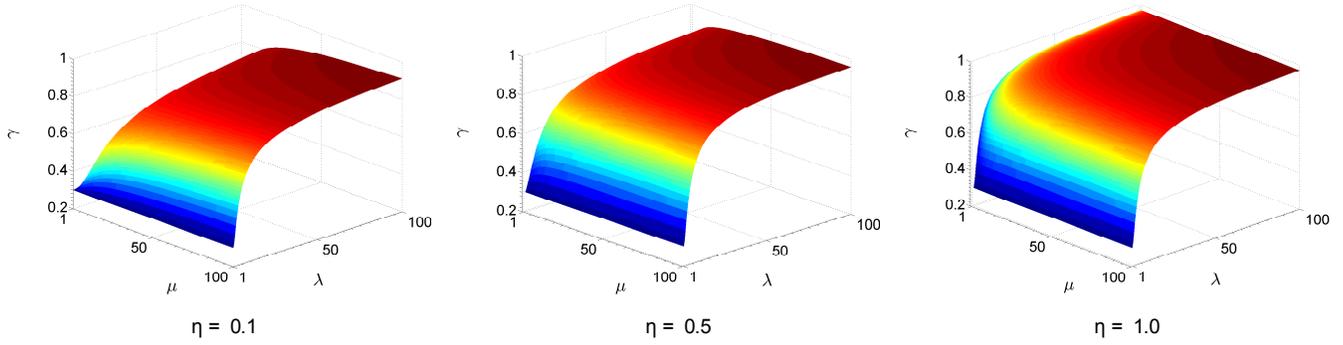}
\caption{\textbf{Surface plots of the performance ratio for various detector efficiencies.}~Here, Arthurs-Kelly uncertainty is neglected in heterodyne detection. The rate of increase in the ratio $\gamma$ is slightly sensitive
 to the value of $\eta$, with the effective $\lambda$-$\mu$ region in which heterodyne detection
significantly outperforms homodyne detection reduces as $\eta$ increases.}
\label{fig:fig2}
\end{figure*}

\begin{figure}[h!]
\centering
\includegraphics[width=\columnwidth]{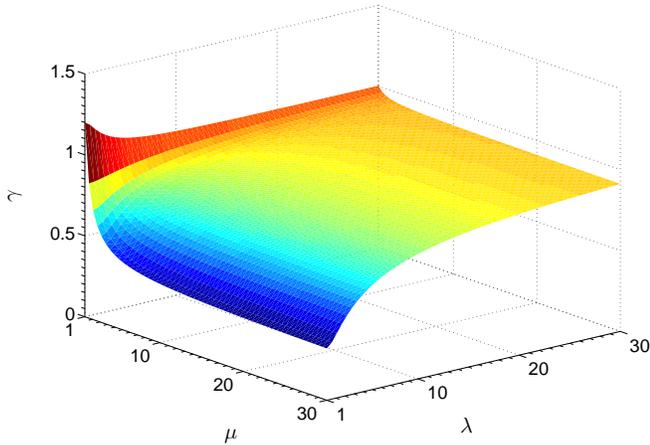}
\caption{\textbf{Surface plots of the performance ratio for perfect detectors.} There exists a small $\lambda$-$\mu$ region within which optimal unbiased covariance-matrix estimators obtained with homodyne detection are more accurate than those obtained with heterodyne. In typical experimental conditions where Gaussian states prepared can neither give rise to minimum uncertainties nor be truely coherent states, there exist a plethora of settings for which $\mathcal{H}_\textsc{het}<\mathcal{H}_\textsc{hom}$.}
\label{fig:fig3}
\end{figure}

\noindent{\textbf{Covariance estimation - revealing the physical consequences.}}~To gain insights in the fundamental difference between
homodyne and heterodyne detection, let us first
 consider the hypothetical situation where there are no quantum-mechanical consequences in
simultaneously measuring two incompatible observables --- the absence of all Arthurs-Kelly-type uncertainties. This entails the equalities
$\delta^{\textsc{(het)}}_\eta=\delta^{\textsc{(hom)}}_\eta=0$
 and $\dyadic{G}_\textsc{hom}=\dyadic{G}_\textsc{het}$ for the detector-efficiency terms and
covariance matrices. The ratio $\gamma=\mathcal{H}_\textsc{het}/\mathcal{H}_\textsc{hom}$ is then
a monotonically increasing function of $\lambda$, $\mu$ and $\eta$. It turns out that this function
has a maximum value of one, which is attained in the limit $\lambda,\mu\rightarrow\infty$.
The ratio $\gamma$ is smallest when $\lambda=\mu=1$, taking the minimal value
of 3/10 for all minimum-uncertainty states ($\mu=1$) with circular
 Wigner-function profiles ($\lambda=1$), \emph{i.e.} the coherent states. Extreme elongation
 of the profiles as a result of huge photonic quadrature squeezing ($\lambda\gg1$) renders both
 CV schemes equivalent in tomographic performance since the significant
 regions of sampling approach phase-space lines of infinite length, and details of the two schemes
 in this hypothetical setting are irrelevant within such infinitesimally thin regions.
 Figure~\ref{fig:fig2} illustrates all the observations made.

The physical implication of these findings can be succinctly written as the following message:
 For \emph{all single-mode Gaussian states}, regardless of the detector efficiency, a direct sampling
 of the phase-space Q~function would in principle give more accurate covariance-matrix
 estimators as compared to their counterparts obtained through
 tomogram processing of data that follow the marginalized distributions of the Wigner function. For this
 class of quantum states, this is consistent with an intuition that a direct inversion of a set of
 statistically consistent data points would yield more accurate Gaussian Wigner-function estimation
 than an indirect two-step post-processing procedure on marginally consistent data points
 that incorporates tomogram combination and data inversion.\\

\begin{figure*}[t]
\centering
\includegraphics[width=1\textwidth]{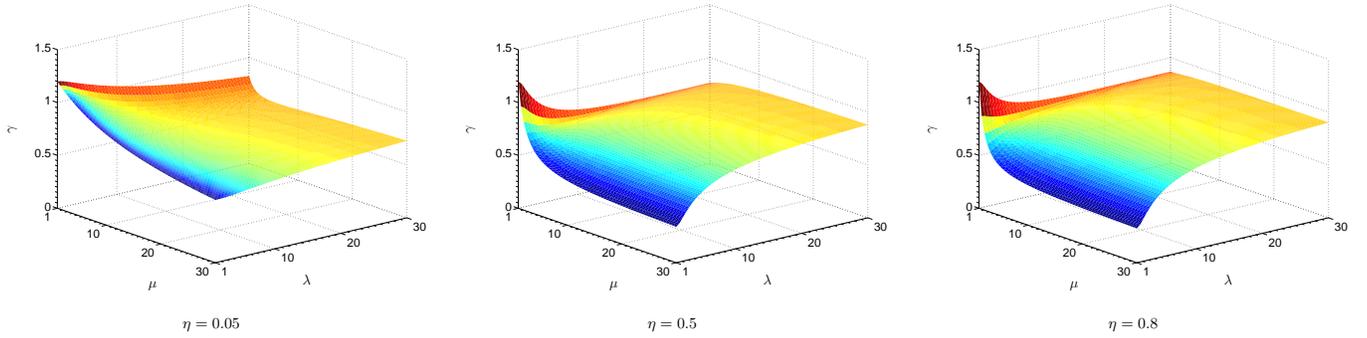}
\caption{\textbf{Surface plots of the performance ratio for various detector efficiencies.} The shapes of surfaces are rather sensitive to the value of $\eta$.}
\label{fig:fig4}
\end{figure*}

\noindent{\textbf{Covariance estimation - analysis for realistic scenarios with perfect detections.}}~In the physical reality, heterodyne detection is always accompanied by the Arthurs-Kelly measurement
uncertainty that comes from simultaneously measuring two complementary quadrature observables.
Physically, the action of the beam splitter (apart from the other two used to carry out
the joint homodyne measurements) that splits the incoming source signal into two in a
heterodyne measurement set-up introduces vacuum fluctuation in the other input photon mode. The result
is an additional vacuum noise that contributes to the overall Arthurs-Kelly-type measurement error
induced by such a measurement, whose lower limit is known to be larger than the usual Heisenberg's
uncertainty lower limit \cite{complem_meas1,complem_meas2}. The formalism of quasi-probability distributions
automatically accounts for the vacuum fluctuation by noting that the covariance matrix
 $\dyadic{G}_\textsc{w}$ for the Gaussian Wigner function of a particular Gaussian state
(associated with homodyne detection) is always less than that for the Q~function,
 $\dyadic{G}_\textsc{q}=\dyadic{G}_\textsc{w}+1/(2\eta)$, by a multiple of the identity that takes
 the usual one-half value when $\eta=1$ for perfect detections.
This additional beam splitter thus plays
the fundamental role in introducing the Arthurs-Kelly uncertainty upon a subsequent measurement.
 It is the combination of this additive vacuum term and the details in handling different types of data
that dictates the executive difference between homodyne and heterodyne detection in state estimation
through the respective covariance matrices $\dyadic{G}_\textsc{hom}$ and $\dyadic{G}_\textsc{het}$
in Eq.~\eqref{eq:hom_het}.

To analyze the consequence of this quantum-mechanical uncertainty, we use the correct expressions for
$\dyadic{G}_\textsc{hom}$ and $\dyadic{G}_\textsc{het}$ and first consider the ideal situation
where the detections are perfect ($\eta=1$). Figure~\ref{fig:fig3} shows the surface plot generated for
Eq.~\eqref{eq:hom_het}. From the plot, we note the maximal influence on $\gamma$ as a manifestion of the Arthurs-Kelly uncertainty for minimum-uncertainty states ($\mu=1$), where $\mathcal{H}_\textsc{het}-\mathcal{H}_\textsc{hom}=1$. The tomographic accuracy associated with heterodyne detection takes the worst-case magnitude $\mathcal{H}_\textsc{het}=6\mathcal{H}_\textsc{hom}/5>\mathcal{H}_\textsc{hom}$ for coherent states and remains greater than unity for all squeezed states ($\lambda>1$). For Gaussian states of higher temperatures ($\mu>1$) that are sufficiently squeezed, $\gamma$ would eventually be smaller than unity, since in the range $\lambda\gg1$, it can be shown that the gradient of $\mathcal{H}_\textsc{hom}$ in $\lambda$ is always steeper than that of $\mathcal{H}_\textsc{het}$. The ratio $\gamma$ approaches unity as $\lambda$ goes to infinity for all $\mu$, in agreement with the previous discussion above.

We conclude from the short excursion above that in this perfect-detection scenario, as long as there exist slight perturbation on a Gaussian source, just as in any realistic setting, such that the resulting quantum state can no longer be of minimum uncertainty, one can always benefit from heterodyne detection with sufficiently large squeezing.\\

\begin{figure}[t]
\centering
\subfigure[~Homodyne detection]{\includegraphics[width=0.45\columnwidth]{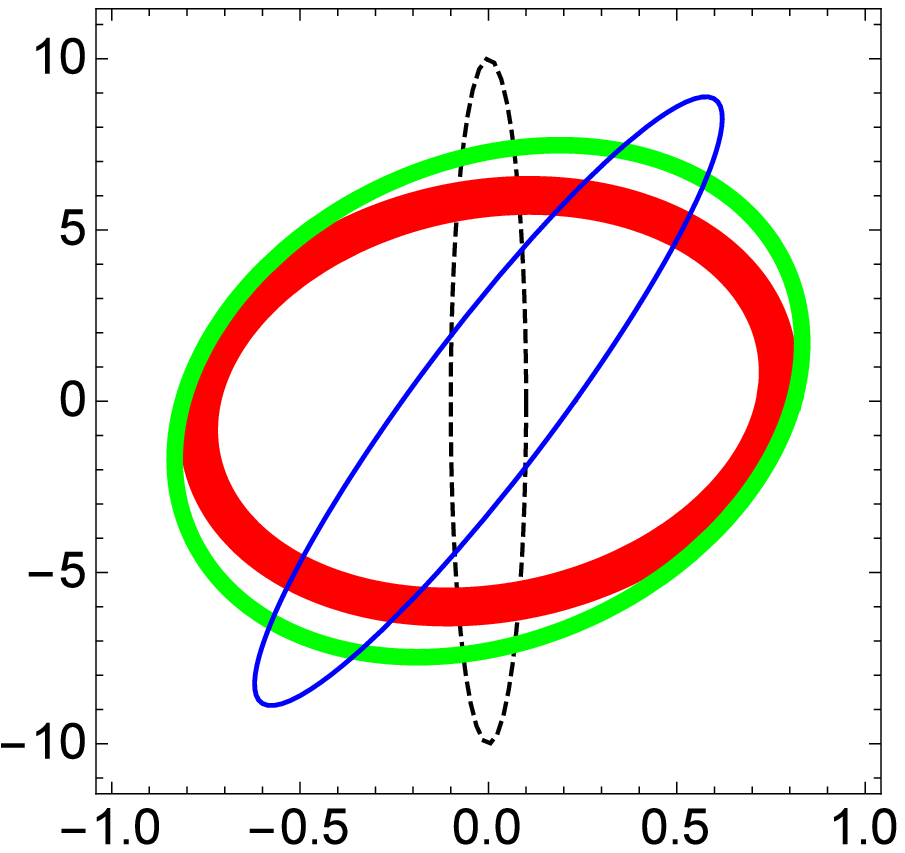}}\qquad
\subfigure[~Heterodyne detection]{\includegraphics[width=0.45\columnwidth]{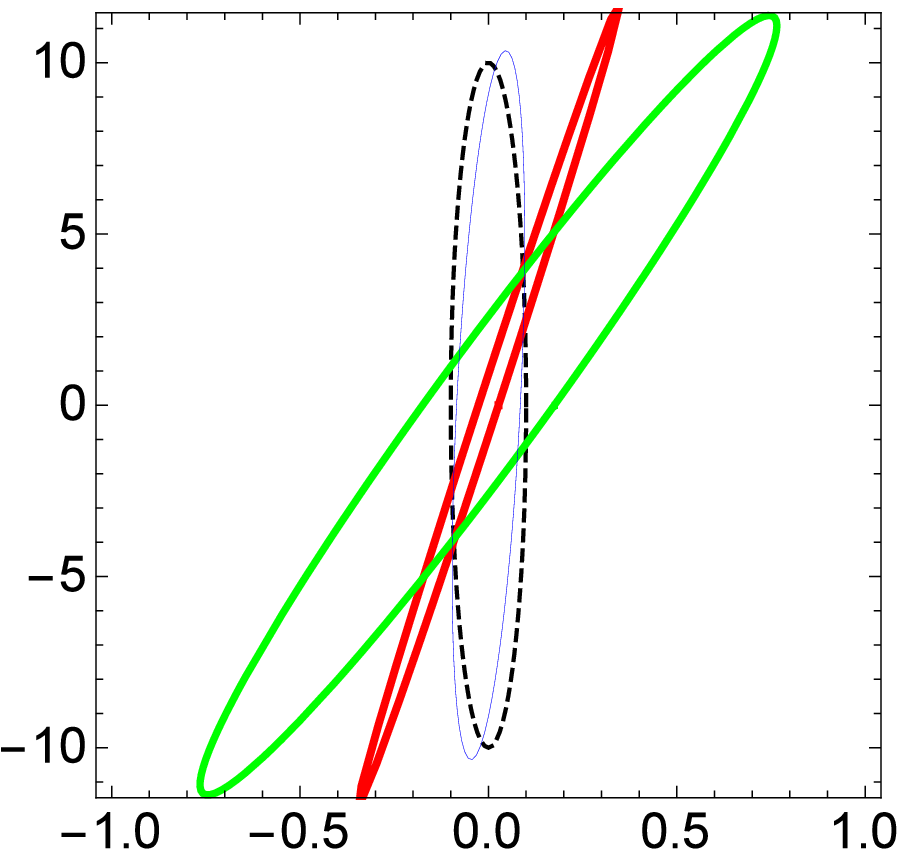}}
\caption{\textbf{Uncertainty-ellipse reconstruction in phase space.} The uncertainty ellipses plotted are reconstructions from particular sets of data that are respectively from measuring $N=50$~(red), 100~(green) and 150~(blue) copies or sampling events in phase space, which are compared to the true uncertainty ellipse (black, dashed). The true Gaussian squeezed state considered here is characterized by the parameters $\mu=2$ and $\lambda=10$, with the detection efficiency $\eta=0.5$. The thickness of each closed curve is proportional to the distance between the reconstructed ellipse and the true ellipse. These plots are representatives of the average performance of the two schemes, a signature of the existence of a wide range of settings for which heterodyne detection is tomographically better than homodyne detection even for moderate values of $N$.}
\label{fig:fig_ellips}
\end{figure}

\noindent{\textbf{Covariance estimation - analysis for realistic scenarios with imperfect detections.}}~In practice, detections are never perfect due to losses, which implies that the detector efficiency $\eta$ is always less than unity. The surface plots for $\gamma$, as shown in Fig.~\ref{fig:fig4}, correspondingly possess rather different shapes and curvatures for different values of $\eta$. The response to $\eta$ for homodyne and heterodyne tomography schemes turn out to be quite different. For example, in the limit of very small efficiency, $\eta\ll 1$, the additional $\eta$-dependent factor $\delta^{\textsc{(het)}}_\eta$ for the heterodyne scheme is almost twice as big as the factor $\delta^{\textsc{(hom)}}_\eta$ for the homodyne scheme. One should expect homodyne tomography to perform better in this limit. Notice that for realistic detection efficiencies in the range $0.2<\eta<0.8$, heterodyne tomography always outperforms homodyne tomography except for a small range of $\mu$ and $\lambda$ parameters. Figure~\ref{fig:fig_ellips} illustrates the performance of the two schemes in terms of uncertainty-ellipse reconstructions of a Gaussian source. Details of its generation are given in \textbf{Methods}.

Numerical calculations show that in this $\eta$-range, the critical values of $\mu$ and $\lambda$ beyond which $\gamma<1$ are respectively $\approx1.736$ and $\approx3.771$. These values correspond to a quadrature squeezing of about $-3.369$~dB and an anti-quadrature squeezing of about $8.1601$~dB from the shot-noise (vacuum) level, which is well below the state-of-the-art squeezing/anti-squeezing levels, which are respectively $\approx-12.7$~dB and $\approx19.9$~dB \cite{qmetro1}. Thus, physicists designing narrowband squeezed Gaussian states with error-suppression of such ratios can enjoy the benefit of tomographically accurate covariance-matrix estimators with heterodyne detection. However, decreasing $\eta$ further to very small values would push the surface up to a level of $6/5$ (the ultimate limit of $\gamma$ for single-mode Gaussian states) eventually, since in the small-$\eta$ limit, the ratio of the asymptotic value of $\mathcal{H}_\textsc{het}\approx6/\eta^2$ to that of $\mathcal{H}_\textsc{hom}\approx5/\eta^2$ becomes a constant.

\bigskip

\noindent
\normalsize{\textbf{Discussion}}\\
\small
Quantum homodyne and heterodyne (double-homodyne) techniques are well-known quantum diagnostic schemes that are accessible in quantum-optics laboratories. The respective tomographic capabilities of these two continuous-variable schemes involve an intricate combination of both the statistical and the data-processing characteristics in each of the schemes. We revealed these inherent tomographic capabilities of heterodyne detection by comparing it with the more popular homodyne detection and analyzing the optimal tomographic accuracy of unbiased covariance-matrix estimation of single-mode Gaussian states with the help of the scaled Fisher information matrix.

Despite the existence of Arthurs-Kelly uncertainties as a consequence of simultaneous measurements of incompatible observables, heterodyne detection exhibits significantly better tomographic performance when probing sufficiently squeezed Gaussian states that are not of minimum-uncertainty, which is the case in all experiments, especially when the detection efficiency is within the practical range in quantum-optics experiments. The amount of quadrature squeezing considered here can be observed with present-day technology.

The seemingly counterintuitive fact that quadrature squeezing can improve heterodyne tomography over homodyne tomography by effectively mask the detrimental effects of intrinsic quantum measurement uncertainties is an important physical consequence that can be easily overlooked if one bases his or her understanding of optimal tomographic performance solely on the manifestation of these uncertainties. The intricate amalgam of statistical and data-processing characteristics in a highly sophisticated fashion is what determines the ultimate tomographic limit for these schemes. This interesting and important revelation justifies the relevance of future experimental work with quantum heterodyne detections to reveal possible novel tomographic enhancements. While we acknowledge the current technical complexities in carrying out heterodyne detection, on which we are in no position to comment, we hope that the conclusions drawn in this article would kick-start some interesting prototypical experimental work.

\bigskip

\noindent
\normalsize{\textbf{Methods}}\\
\scriptsize
\noindent{\textbf{Uncertainty regions.}}~To analyze the uncertainty regions more thoroughly, let us focus on the uncertainty regions of the two tomography schemes. These regions are centered at the origin of the phase space, and in terms of the angle $\theta$, their respective boundaries are described by the functions $\sigma_\theta$ and $\Sigma_\theta$. These are precisely the standard deviations for the two schemes of a fixed angle $\theta$. For simplicity we
assume ideal detector efficiency $\eta=1$. Then, with Gauss-Weierstrass transform that turns Wigner functions into Q functions, and \emph{vice versa}, the respective covariance matrices $\dyadic{G}_\textsc{w}$ and $\dyadic{G}_\textsc{q}$ are related by the expression $\dyadic{G}_\textsc{q}=\dyadic{G}_\textsc{w}+\dyadic{1}/2$.

For minimum uncertainty states, where
\begin{equation}
\dyadic{G}_\textsc{w}\,\widehat{=}\left(
\begin{array}{cc}
\dfrac{1}{2\lambda} & 0\\
0 & \dfrac{\lambda}{2}
\end{array}
\right)\,,
\end{equation}
we have
\begin{equation}
\sigma_\theta=\sqrt{\frac{1}{2\lambda}(\cos\theta)^2+\frac{\lambda}{2}(\sin\theta)^2}
\end{equation}
and
\begin{equation}
\Sigma_\theta=\left[\sqrt{1+\dfrac{\lambda-1}{\lambda+1}\cos(2\theta)}\right]^{-1}\,,
\label{eq:Sig2_Q}
\end{equation}
whence we immediately recognize that Eq.~\eqref{eq:Sig2_Q} is the equation of an ellipse in polar coordinates. This ellipse marks the contour line
of constant 2-D Q function. The uncertainty region for homodyne tomography is more complicated, but for states of highly elongated regions, that is $\lambda\gg 1$ or $\lambda \ll 1$, this region looks like two circles in contact. Plots of these uncertainty regions are given below.

\begin{figure}[h!]
\centering
\subfigure[\scriptsize\,\,$\lambda=1/2$]{\includegraphics[width=0.45\columnwidth]{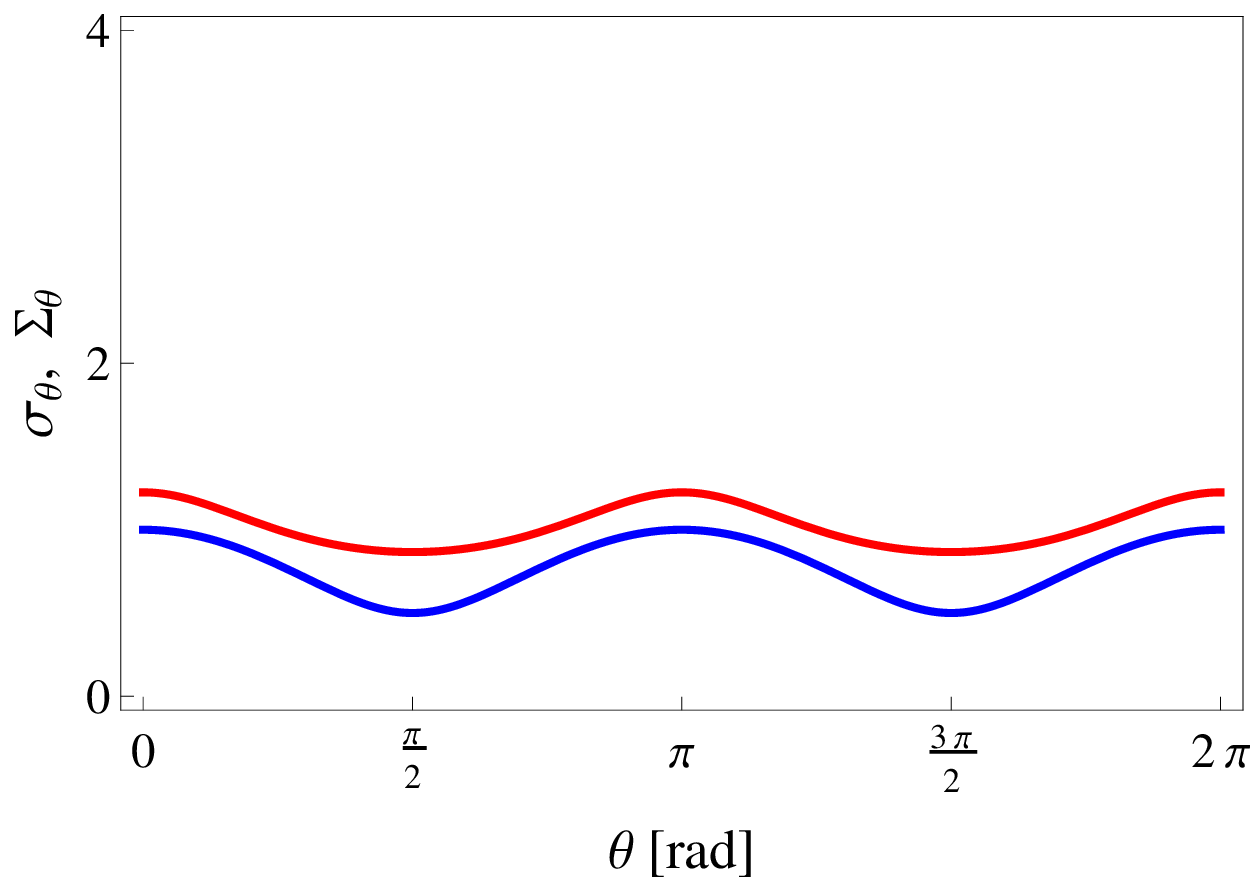}}
\subfigure[\scriptsize\,\,$\lambda=1/8$]{\includegraphics[width=0.45\columnwidth]{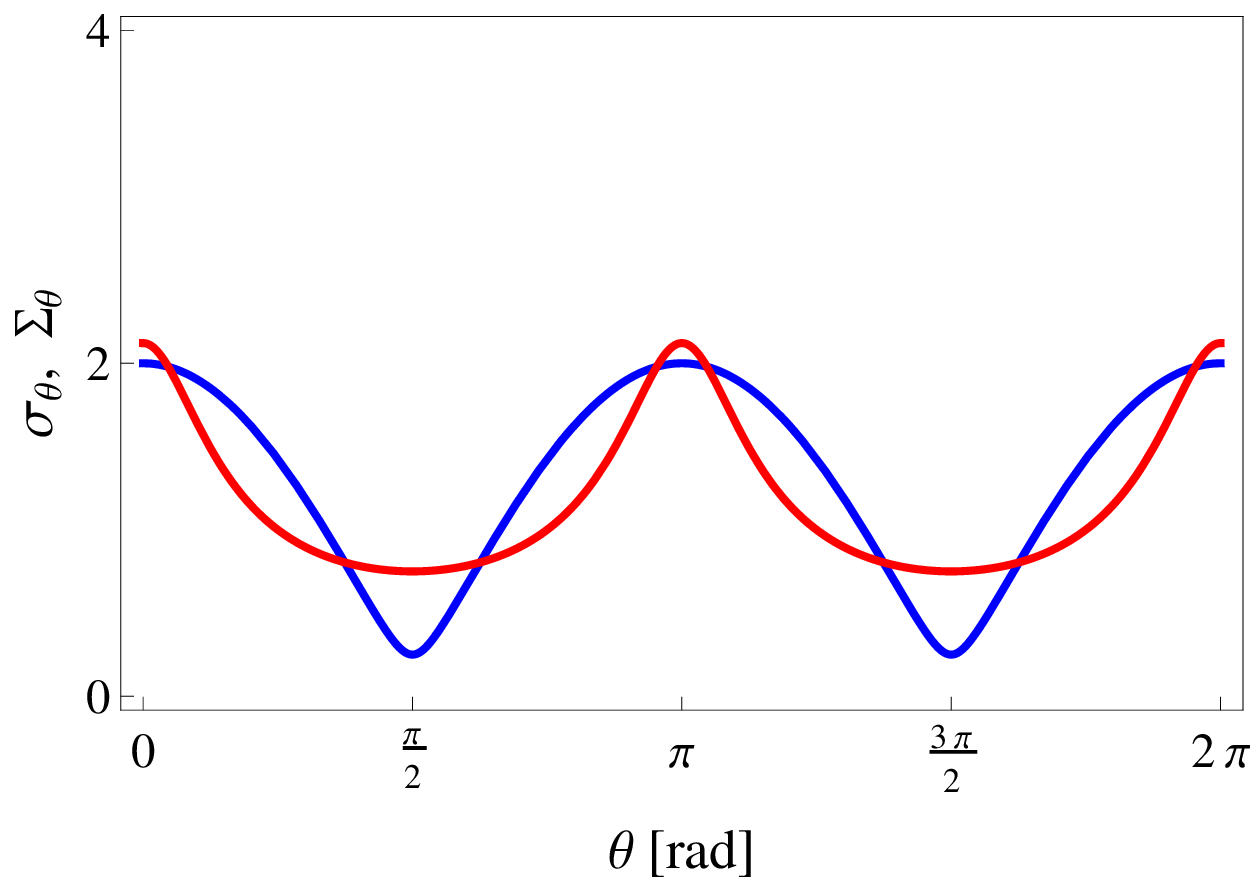}}\\
\subfigure[\scriptsize\,\,$\lambda=1/32$]{\includegraphics[width=0.45\columnwidth]{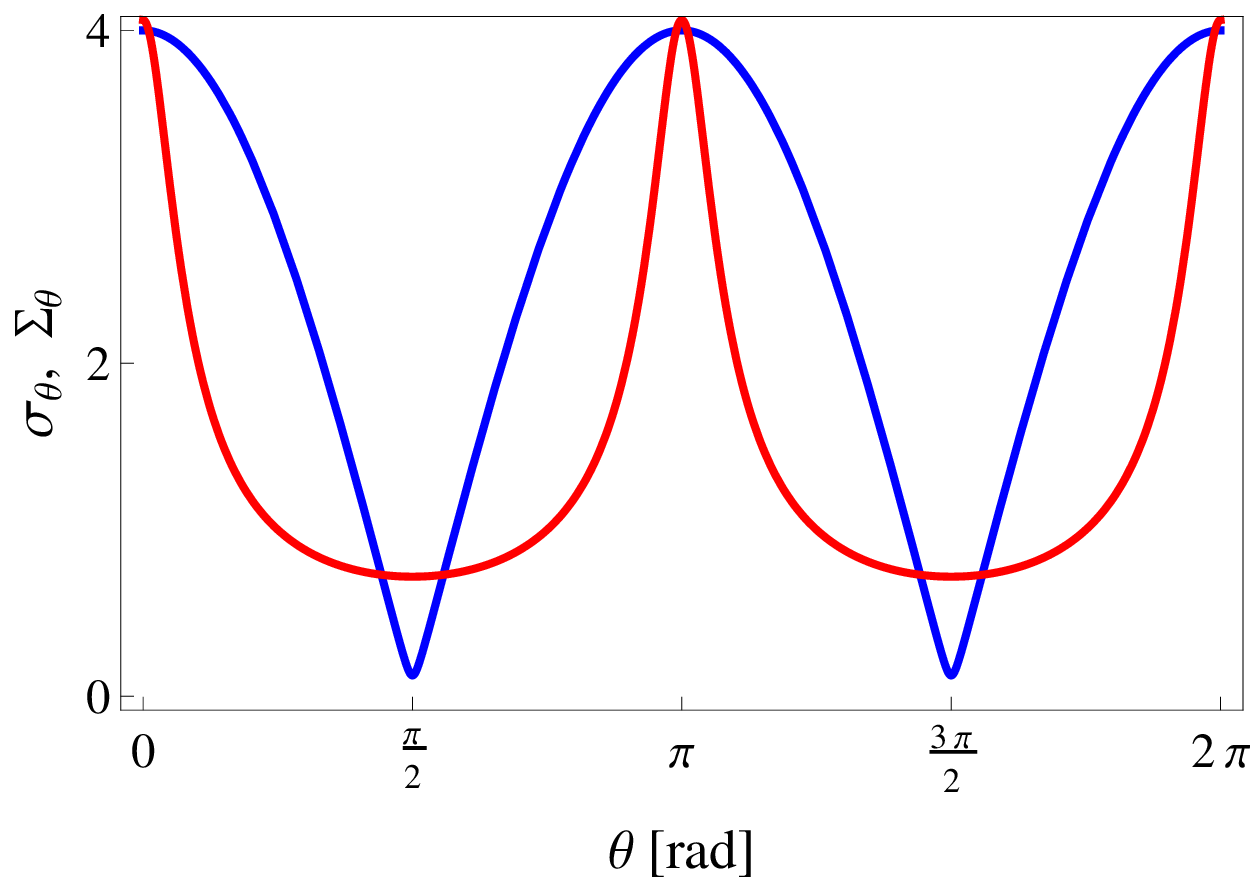}}
\caption{\scriptsize\textbf{Plots of $\sigma_\theta$ (Blue) and $\Sigma_\theta$ (Red) as functions of $\theta$.}}
\label{fig:sig_Sig1}
\end{figure}

\begin{figure}[h!]
\centering
\subfigure[\scriptsize\,\,$\lambda=1$]{\includegraphics[width=0.45\columnwidth]{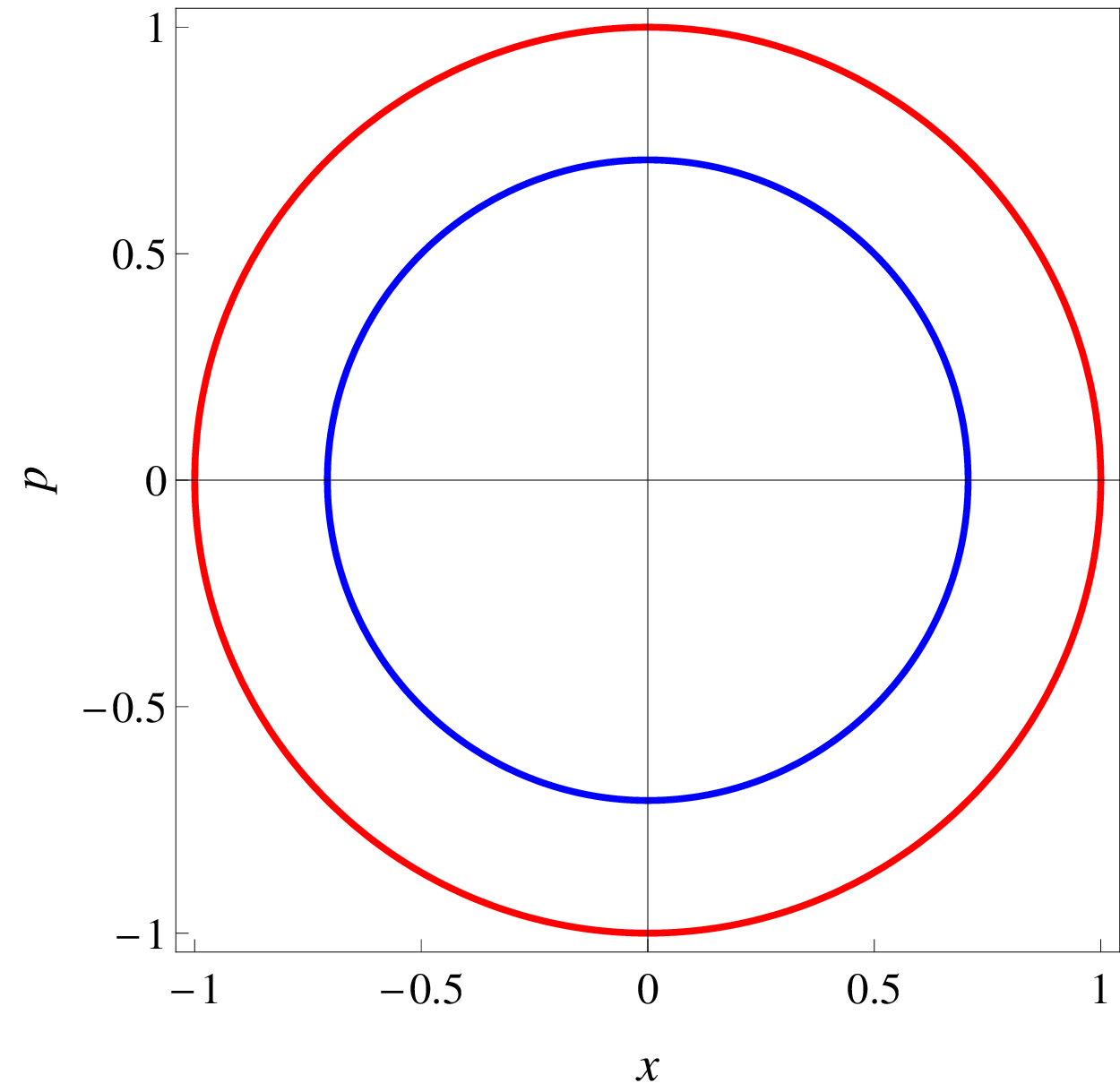}}
\subfigure[\scriptsize\,\,$\lambda=1/4$]{\includegraphics[width=0.45\columnwidth]{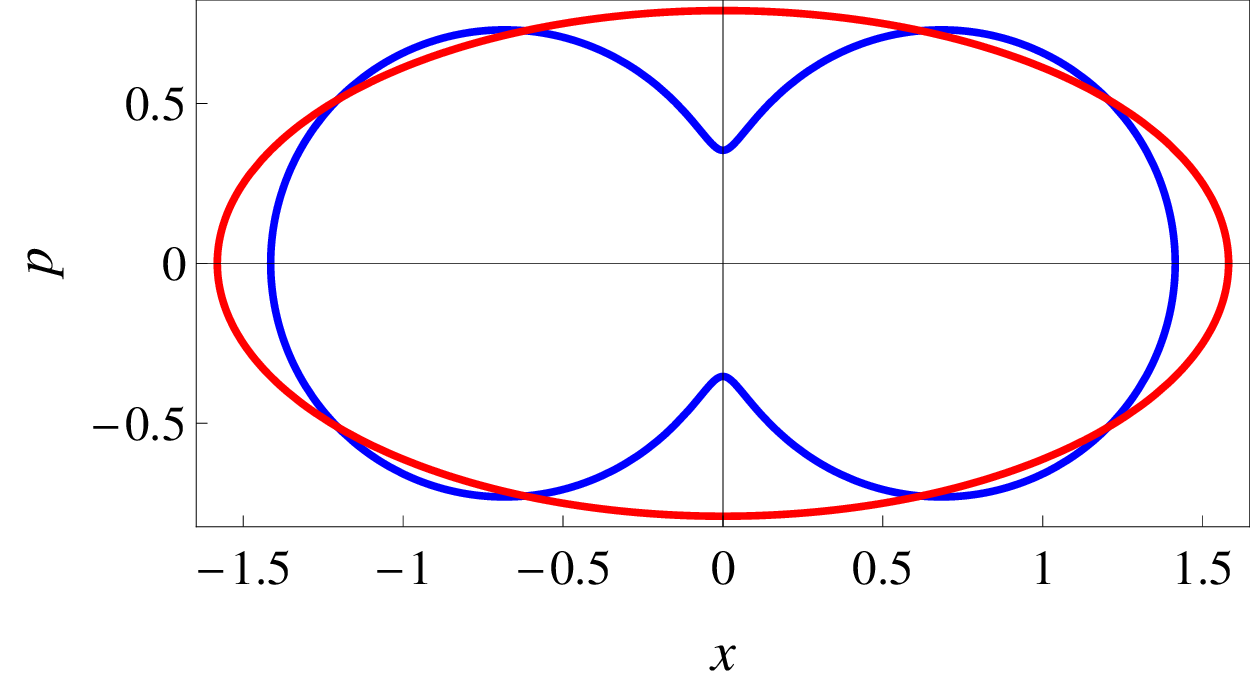}}\\
\subfigure[\scriptsize\,\,$\lambda=1/16$]{\includegraphics[width=0.45\columnwidth]{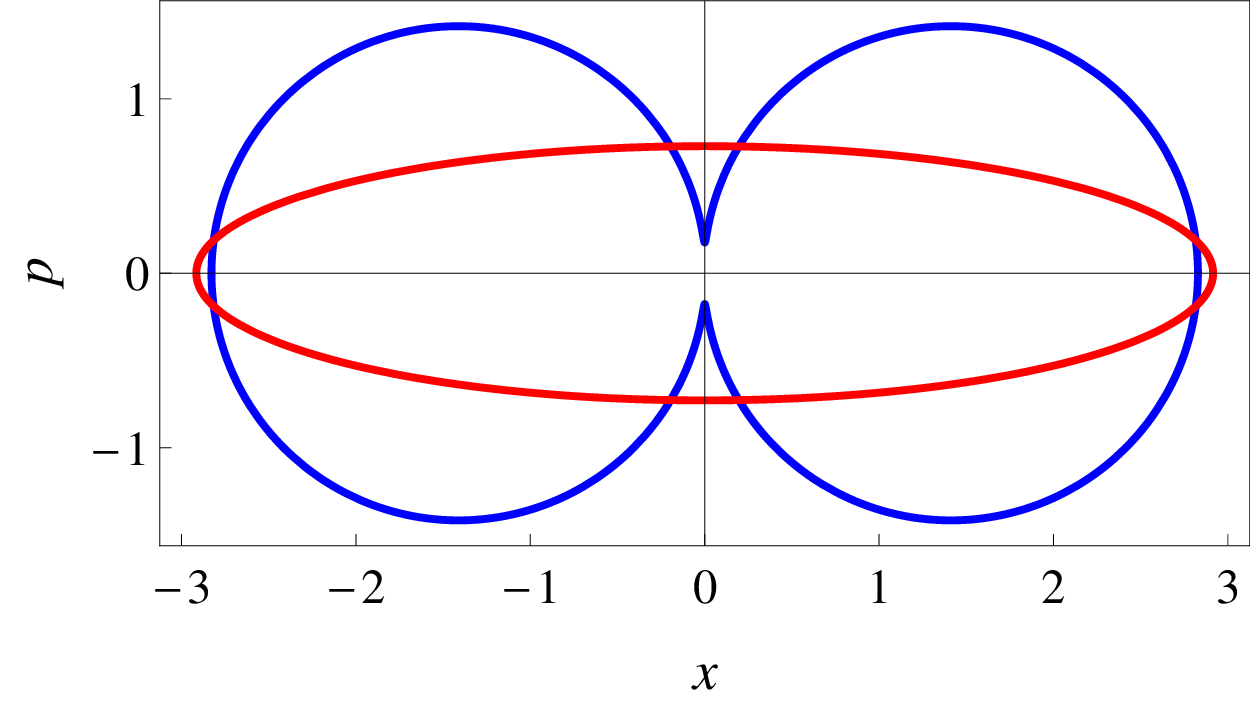}}
\subfigure[\scriptsize\,\,$\lambda=1/128$]{\includegraphics[width=0.45\columnwidth]{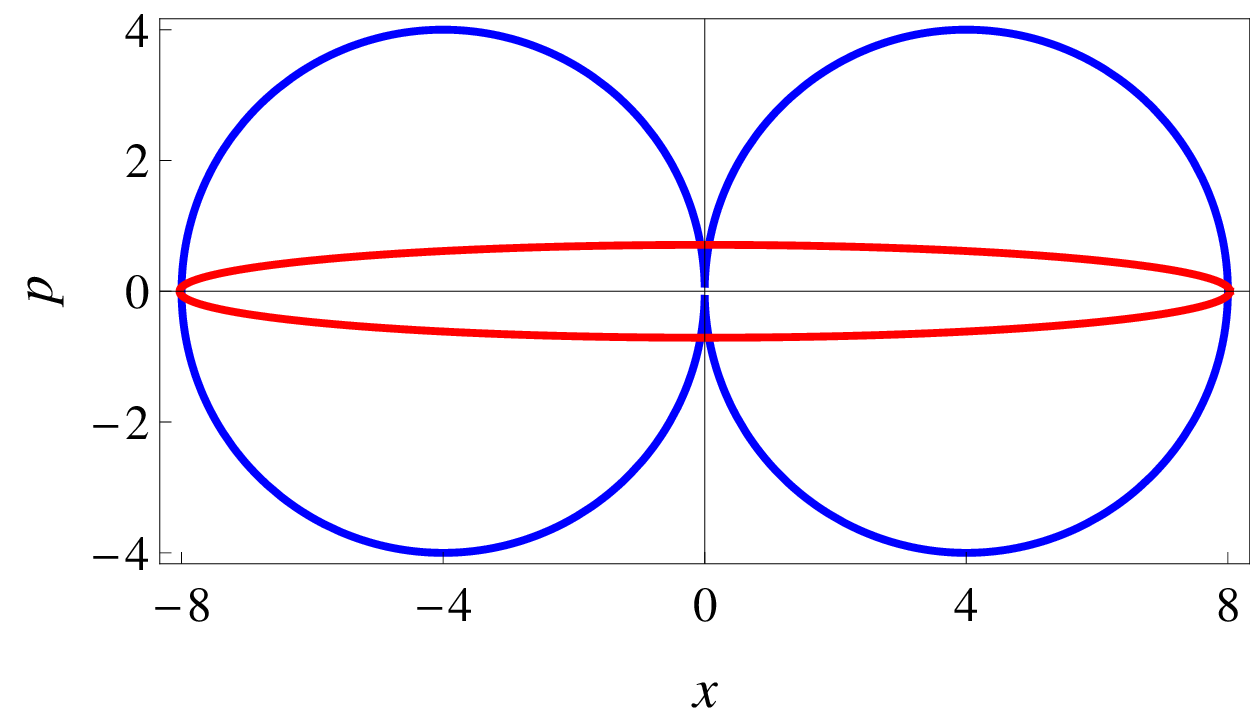}}
\caption{\scriptsize\textbf{Uncertainty regions for homodyne (Blue) and heterodyne (Red) tomography in phase space.}}
\label{fig:sig_Sig2}
\end{figure}

Figure~\ref{fig:sig_Sig1} shows that homodyne detection seems to give smaller errors for nearly circular states, while for states of highly elongated Wigner functions, uncertainty suppression is observed with heterodyne detection for almost all angles. Naturally, if we integrate/cut approximately along the direction of a principal axis, homodyne data is less noisy. However, the interval of angles for which $\sigma_\theta<\Sigma_\theta$ shrinks as the Wigner function becomes more elongated. Figure~\ref{fig:sig_Sig2} shows the uncertainty regions for the two schemes.

\noindent
The area of the heterodyne uncertainty ellipse is given by
\begin{equation}
S_\Sigma=\pi\,\sqrt{\frac{1}{2\lambda}+\frac{1}{2}}\sqrt{\frac{\lambda}{2}+\frac{1}{2}}
=\pi \frac{\lambda+1}{2\sqrt{\lambda}}\,,
\end{equation}
whereas that of the quartic homodyne uncertainty region is
\begin{equation}
S_\sigma=\pi\,\frac{\lambda^2+1}{4\lambda}.
\end{equation}
The latter is $\sqrt{\lambda}/2$ $\left[1/(2\sqrt{\lambda})\right]$ times more than the former in the limit of
$\lambda\gg 1$ [$\lambda\ll 1$].
The areas become equal when
\begin{equation}
\lambda=\lambda_{\text{crit}}=1+\sqrt{3}\pm\sqrt{3+2\sqrt{3}}=
\left\{\begin{array}{l}
5.2745\,,\\
0.18959\,.
\end{array}
\right.
\end{equation}
One solution is the reciprocal of the other as it should be, for the values $\lambda=\lambda_{\text{crit}}$ and $\lambda=1/\lambda_{\text{crit}}$ essentially correspond to the same Gaussian state apart from a rotation by a $\pi/2$ angle.

\begin{figure}[h!]
\centering
\scriptsize
\includegraphics[width=0.8\columnwidth]{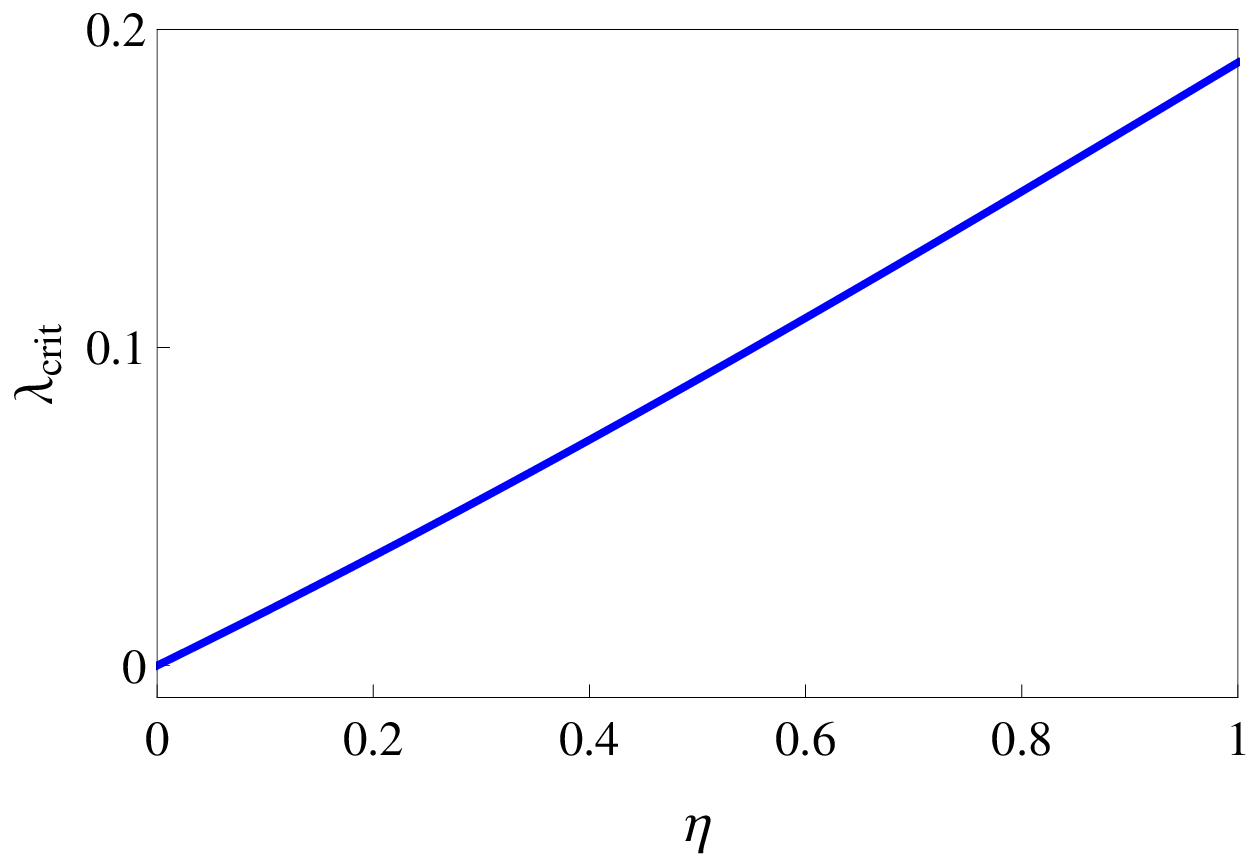}
\label{fig:equal}
\caption{\scriptsize\textbf{Plot of $\lambda_\text{crit}$ against $\eta$ for which $S_\sigma=S_\Sigma$.}}
\end{figure}

Generalization of the above to inefficient detections --- $\eta<1$ --- is possible and closed form
expressions for the uncertainty areas are available. Figure~\ref{fig:equal} shows the plot of the critical value $\lambda_\text{crit}$ for which the homodyne and heterodyne areas are equal as a function of the detector efficiency $\eta$.

Smaller detection efficiencies mean larger differences between the corresponding Wigner and Q functions and
more state asymmetry (larger $\lambda_\text{crit}$ in other words) is needed to see the effect. A realistic choice of $\eta=0.8$ gives $\lambda=0.149$ which requires the smallest standard deviation to be squeezed by a factor of $2.6$ below the shot noise, well within the experimentally feasible range for optical quadrature squeezing.

Needless to say, the minimum-uncertainty states discussed above are the worst possible states for this kind of comparison since the extra noise from the Q~function stands out. With a lot of excess noise (mixed states), as we always have in actual experiments, where the anti-squeezed quadrature is never controlled as well as the squeezed one, this extra vacuum term matters less and the uncertainty suppression in heterodyne tomography over homodyne tomography will be easy to observe.\\

\noindent{\textbf{Covariance estimation --- background.}}~We are interested in estimating the true covariance matrix of the Wigner function that is parametrized as
\begin{equation}
\label{parametrization}
\dyadic{G}_\textsc{w}\,\widehat{=}\left(
\begin{array}{cc}
g_1 & g_3/\sqrt{2}\\
g_3/\sqrt{2} & g_2
\end{array}
\right),
\end{equation}
so that this covariance matrix $\dyadic{G}_\textsc{w}=\sum_{k=1}^3 g_k \dyadic{\Gamma}_k$ can be written as a linear combination of trace-orthonormal Hermitian basis matrices $\dyadic{\Gamma}_k$ --- $\SP{\dyadic{\Gamma}_k\dyadic{\Gamma}_l}=\delta_{kl}$.

Let $\widehat{\dyadic{G}}_\textsc{w}$ be an estimator of $\dyadic{G}_\textsc{w}$. We quantify the performance of this estimator using the scaled mean squared-error (MSE) with $N$ that is defined to be the mean Hilbert-Schmidt distance between the true covariance matrix and the estimator,
\begin{equation}
\label{eq:mse}
\mathcal{H}=\overline{\left(\dyadic{G}_\textsc{w}-\widehat{\dyadic{G}}_\textsc{w}\right)^2}=\sum_k \overline{\left(g_k-\widehat{g}_k\right)^2}.
\end{equation}
There are two kinds of estimators that are popular: the linear-inversion (LIN) estimator and maximum-likelihood (ML) estimator. Only the ML estimator is guaranteed to be positive. For our purposes, we shall consider the limit of large $N$, where both estimators coincide and are unbiased. For these estimators, the scaled MSE $H$ is bounded from below according to the inequality
\begin{equation}
\label{crlb}
\mathcal{H}\geq\SP{\dyadic{F}^{-1}}\,,
\end{equation}
where $\dyadic{F}$ is the scaled Fisher information matrix that is represented by a $3\times3$ positive real matrix. The right-hand side of \eqref{crlb} is known as the Cram{\'e}r-Rao bound. By denoting the row of parameters as $\TP{\rvec{g}}\,\widehat{=}\,(g_1\,\,\,g_2\,\,\,g_3)$, for multivariate Gaussian statistics described by the covariance matrix $\dyadic{G}$, the scaled Fisher matrix reads
\begin{equation}
\label{fisher}
\dyadic{F}=\frac{1}{2}\SP{\dyadic{G}^{-1} \frac{\partial \dyadic{G}}{\partial \rvec{g}} \dyadic{G}^{-1}
\frac{\partial \dyadic{G}}{\partial \rvec{g}}}\,,
\end{equation}
where we remind the reader that the matrix trace acts only on the $\dyadic{G}$s. The derivation of this result is simply a slight generalization of that employed in Ref.~\cite{fisher-gauss}.\\

\noindent{\textbf{Covariance estimation --- homodyne tomography.}}~In this case, we are sampling from marginal distributions. Different quadratures contribute to the total scaled Fisher matrix independently. This is easy to see. Out of $d$ different quadratures, drawing one sample per quadrature amounts to drawing a $d$-dimensional random sample from the $d$-dimensional multivariate Gaussian distribution described by a diagonal covariance matrix $\dyadic{C}$ with entries
\begin{equation}
\label{Gii}
C_{jj}\equiv C(\theta_j)=\TP{\rvec{u}}_{\theta_j}\,\dyadic{G}_\textsc{hom}\,\rvec{u}_{\theta_j}, \qquad j=1,\ldots,d,
\end{equation}
where
\begin{align}
\dyadic{G}_\textsc{hom}&=\dyadic{G}_\textsc{w}+\delta^{\textsc{(hom)}}_\eta\dyadic{1}\,,\nonumber\\
\delta^{\textsc{(hom)}}_\eta&=\dfrac{1-\eta}{2\eta}\,.
\end{align}
The angle $\theta_j$ defines the $j$th quadrature angle setting. All off-diagonal terms are zero because measurements of different quadratures are independent and uncorrelated. Using Eq.~\eqref{fisher}, we find
\begin{equation}
\dyadic{F}=\frac{1}{d}\sum_{j=1}^d \dyadic{f}(\theta_j),
\end{equation}
where
\begin{equation}
\dyadic{f}(\theta)=\frac{1}{2C(\theta)^2}  \frac{\partial C(\theta)}{\partial \rvec{g}}
\frac{\partial C(\theta)}{\partial \rvec{g}}
\end{equation}
is the scaled Fisher matrix of a single quadrature, which is really a special case of Eq.~\eqref{fisher}, and the normalization by $d$ is there to account for splitting the total ensemble among $d$ quadratures.
Finally, for a fair comparison in phase space, we take the limit of infinitely many quadrature angle settings and replace the sum by the integral. This also closely follows the experimental practice, where rather
than measuring a fixed number of quadratures, the phase of the local oscillator is continuously being
changed during the experiment, so that effectively a very large number of quadratures (up to $10^6$)
is realized. Thereafter,
\begin{equation}
\dyadic{F}=\int_0^\pi\,\dfrac{\D\theta}{\pi}\,\dyadic{f}(\theta).
\label{eq:fisher_hom}
\end{equation}

We calculate the scaled Fisher matrix and scaled MSE for both homodyne and heterodyne schemes. The parametrization of Eq.~\eqref{parametrization} is used. The easiest way to do this is to set $g_3=0$, and after which, employ the invariance properties of the scaled Fisher matrix. Once the final scaled MSE formulas are cast in basis-independent form, that is, in terms of matrix invariants, they become general.

Indeed, in heterodyne detection, any $2\times2$ covariance matrix can be diagonalized by applying a rotation $\mathbf{R}$, according to which a new
trace-orthonormal operator basis is generated: $\dyadic{\Gamma}_k \rightarrow \dyadic{\Gamma}_k'=\dyadic{R}\, \dyadic{\Gamma}_k \,\dyadic{R}^T$,  $\dyadic{R} \,\dyadic{R}^T=\dyadic{R}^T \,\dyadic{R}=\dyadic{1}$ and $\SP{\dyadic{\Gamma}_k' \dyadic{\Gamma}_l'}=\delta_{kl}$. Expressing the old basis in terms of the new basis elements, $\dyadic{\Gamma}_k=\sum_l w_{kl}  \dyadic{\Gamma}_l'$, the scaled Fisher matrix transforms into $\dyadic{F}\rightarrow \dyadic{W}^T\,\dyadic{F}\,\dyadic{W}$, where orthonormality of the new basis implies that $\dyadic{W}\dyadic{W}^T=\dyadic{W}^T\dyadic{W}=\dyadic{1}$. Therefore, the scaled MSE, which is given by the matrix trace of $\dyadic{F}^{-1}$, does not change. Alternatively, one can rotate the covariance matrix, rather than the coordinate system, and see that the right-hand side of Eq.~\eqref{fisher} does not change upon the mapping $\dyadic{G} \rightarrow \dyadic{R}^T\, \dyadic{G} \,\dyadic{R}$.

In the case of homodyne detection, the invariance with respect to a change in the orientation of the uncertainty ellipse follows from integrating the scaled Fisher information over all quadratures (angles).

For homodyne measurements over the \emph{entire} phase space, there is, unfortunately, no closed-form expression for the LIN/ML estimator, even for large $N$. However, one can still compute the scaled MSE in the large-$N$ limit from the scaled Fisher matrix in Eq.~\eqref{eq:fisher_hom}. This requires calculating integrals over all angles between zero and $\pi$, which can be carried out easily using contour integration techniques.

By defining
\begin{equation}
\label{formgam}
\beta\equiv\dfrac{\SP{\dyadic{G}_\textsc{hom}}+2\sqrt{\DET{\dyadic{G}_\textsc{hom}}}}{g_1-g_2}\,,
\end{equation}
the expression for the scaled Fisher matrix ($g_3=0$) reads
\begin{equation}
\dyadic{F}_\textsc{hom}\,\widehat{=}\,\dfrac{1}{(g_1-g_2)^2}\begin{pmatrix}
\dfrac{1+3\beta}{(1+\beta)^3} & \dfrac{1}{\beta^2-1} & 0\\[0.35cm]
\dfrac{1}{\beta^2-1} & \dfrac{1-3\beta}{(1-\beta)^3} & 0\\[0.35cm]
0 & 0 & \dfrac{2}{\beta^2-1}
\end{pmatrix}\,.
\end{equation}
With this, the Cram{\'e}r-Rao bound is given by
\begin{equation}
\label{hhom}
\mathcal{H}_\textsc{hom}=\dfrac{(g_1-g_2)^2}{4\beta^2}\left(5\beta^4+4\beta^2-1\right)\,.
\end{equation}
This formula turns into the more general form after replacing $g_1-g_2$ in Eq.~\eqref{hhom}
by the difference in eigenvalues of $\dyadic{G}_\textsc{hom}$, thus yielding the first equation of \eqref{eq:hom_het}.\\

\noindent{\textbf{Covariance estimation --- heterodyne tomography.}}~The relevant covariance matrix is
\begin{align}
\dyadic{G}_\textsc{het}&=\dyadic{G}_\textsc{w}+\delta^{\textsc{(het)}}_\eta\dyadic{1}\,,\nonumber\\
\delta^{\textsc{(het)}}_\eta&=\dfrac{2-\eta}{2\eta}\,.
\end{align}
Since in heterodyne detection, we are \emph{directly} sampling the 2-D multivariate Gaussian distribution, subtracting the $\eta$-dependent term $\delta^{\textsc{(het)}}_\eta$ from the sample covariance matrix gives the efficient estimator that attains the Cram{\'e}r-Rao bound.

For this type of tomography, the optimal LIN estimator is well-known. It is essentially given by the total sample covariance matrix of all the collected data two-tuples $\{(x_j,p_j)\}$, up to an additive $\eta$-dependent multiple of the identity. The scaled MSE of this LIN estimator (also of the ML estimator for large $N$) can easily be computed either by calculating the scaled Fisher matrix ($g_3=0$) using Eq.~\eqref{fisher}
\begin{equation}
\dyadic{F}_\textsc{het}\,\widehat{=}\,\dfrac{1}{2}\begin{pmatrix}
\dfrac{1}{[(\dyadic{G}_\textsc{het})_{11}]^{2}} & 0 & 0\\[0.35cm]
0 & \dfrac{1}{[(\dyadic{G}_\textsc{het})_{22}]^2} & 0\\[0.35cm]
0 & 0 & \dfrac{1}{(\dyadic{G}_\textsc{het})_{11} (\dyadic{G}_\textsc{het})_{22}}
\end{pmatrix}\,
\end{equation}
and then take its inverse, or by directly performing the average over all data of the Hilbert-Schmidt distance in Eq.~\eqref{eq:mse} under the consideration of 2-D Gaussian statistics. The two approaches are equivalent since direct-sampling of the Q function results in 2-D Gaussian statistics. Either way, the closed-form expression, written in the basis-independent form, is the one given as the second equation of \eqref{eq:hom_het}.\\

\noindent{\textbf{Covariance estimation --- Comparisons of reconstructed uncertainty ellipses.}}~Figure~\ref{fig:fig_ellips} compares reconstructed uncertainty ellipses for a given true squeezed Gaussian state using the two CV schemes. The parameters for this state are chosen so that $\gamma<1$ in the large $N$ limit. Here, the potential of heterodyne detection can be witnessed even for moderate values of $N$.

To generate the curves, we investigate unbiased maximum-likelihood (ML) covariance-matrix estimators $\widehat{\GMAT}_\textsc{ml}$ that are in principle asymptotically optimal with respect to the Cram{\'e}r-Rao bound for the Hilbert-Schmidt distance $\tr{\Big(\widehat{\GMAT}_\textsc{ml}-\GMAT_\text{true}\Big)^2}$ which we take as the measure for tomographic accuracy. The ML estimator for the homodyne data is obtained by simply taking all data obtained from various angles and performing a maximization of the likelihood, which is a Gaussian distribution function characterized by elements of the covariance matrix, to obtain the covariance matrix that maximizes this likelihood. To obtain the ML estimator for the heterodyne data, the sampled phase-space points are gathered and the corresponding sample covariance matrix is calculated from these points. This sample covariance matrix is, by definition, the maximum-likelihood estimator for Gaussian states.

From these covariance matrices, uncertainty ellipses can be directly obtained by computing the eigenvalues and eigenvectors of these matrices, where the two orthonormal eigenvectors of each matrix represent the basis vectors for the two principle axes, and the eigenvalues represent the lengths of these axes.

\bigskip

\noindent
\normalsize{\textbf{Acknowledgments}}\\
\small
This work is co-financed by the European Social Fund and the state budget of the Czech Republic, project~No.~CZ.1.07/2.3.00/30.0004 (POST-UP), and partially supported by both the IGA Project of the Palacky University (Grant No. PRF~2014-014) and the European Union Seventh Framework Programme under Grant Agreement No. 308803 (Project BRISQ2).\\

\bigskip

\noindent
\normalsize{\textbf{Contributions}}\\
\small
The project was initiated by Z.~H. and J.~R.. Analytical and numerical calculations were done by Y.~S.~T. and J.~R.. Z.~H., J.~R. and S.~W. jointly supervised this project. This manuscript was written by Y.~S.~T., and further inputs and suggestions were given by J.~R., Z.~H. and S.~W..

\bigskip

\noindent
\normalsize{\textbf{Additional information}}\\
\small
The author(s) declare no competing interests as defined by Nature Publishing Group, or other interests that might be perceived to influence the results and/or discussion reported in this paper.


\end{document}